\documentclass{Interspeech}

\interspeechcameraready

\title{No Audiogram: Leveraging Existing Scores for Personalized Speech Intelligibility Prediction \thanks{* Corresponding authors}}

\author[affiliation={1}]{Haoshuai}{Zhou}
\author[affiliation={1}]{Changgeng}{Mo}
\author[affiliation={1}]{Boxuan}{Cao}
\author[affiliation={1,2,*}]{Linkai}{Li}
\author[affiliation={2,3,*}]{Shan Xiang}{Wang}

\affiliation{}{Orka Labs Inc.}{China}
\affiliation{Electrical Engineering}{Stanford University}{United States}
\affiliation{Materials Science and Engineering}{Stanford University}{United States}
\email{\{haoshuai,dicky.mo,boxuan.cao\}@hiorka.com, \{linkaili,sxwang\}@stanford.edu}

\usepackage{comment}
\usepackage{multirow}

\begin{document}

\maketitle

\begin{abstract}
    
    Personalized speech intelligibility prediction is challenging. Previous approaches have mainly relied on audiograms, which are inherently limited in accuracy as they only capture a listener's hearing threshold for pure tones. Rather than incorporating additional listener features, we propose a novel approach that leverages an individual's existing intelligibility data to predict their performance on new audio. We introduce the Support Sample-Based Intelligibility Prediction Network (SSIPNet), a deep learning model that leverages speech foundation models to build a high-dimensional representation of a listener's speech recognition ability from multiple support (audio, score) pairs, enabling accurate predictions for unseen audio. Results on the Clarity Prediction Challenge dataset show that, even with a small number of support (audio, score) pairs, our method outperforms audiogram-based predictions. Our work presents a new paradigm for personalized speech intelligibility prediction.
\end{abstract}

\section{Introduction}
Metrics are crucial for algorithm optimization. For perceptual speech algorithms, such as speech enhancement and wide dynamic range compression, speech intelligibility score is one of the most important metrics. However, obtaining accurate speech intelligibility data is non-trivial, as it requires conducting numerous listening tests with real individuals. This process is not only costly but also non-automated, making it impractical for supporting rapid algorithm iteration.

To address this issue, many alternative objective metrics have been proposed \cite{5495701, 10.1109/TASLP.2016.2585878, zezario22_interspeech, pedersen20_interspeech, 10.1109/TASLP.2018.2847459}. However, most of these metrics are designed for normal hearing individuals. Recently, some studies have started focusing on objective speech intelligibility score for hearing-impaired people \cite{Zezario2022MBINetAN, 10447907, robach22_interspeech, 10447597}. The Clarity Prediction Challenge \cite{barker22_interspeech, 10446441} organizes an open competition that enables participants to evaluate intelligibility prediction performance using intelligibility scores and audiogram data collected from hearing-impaired individuals, thereby significantly accelerating progress in this research domain.
 
Although it is common practice to use listener audiograms as additional input for hearing-related information, audiograms only provide a broad measure of hearing thresholds, lacking important cognitive, perceptual, and contextual factors that play a key role in actual speech comprehension \cite{Parmar2021FactorsAT, Roberts2016PerceptionAC, Boothroyd1988MathematicalTO, 10.1007/978-3-319-25474-6_5}. Figure 1 shows that on the CPC dataset, individuals with similar average hearing levels can have significant differences in their speech intelligibility score, which raises questions about the effectiveness of predicting speech intelligibility score based solely on one's audiograms.

Rather than relying on additional features such as frequency selectivity capability \cite{Preminger1985FrequencySA}, language experience \cite{Bradlow1999RecognitionOS} and cognitive ability \cite{Heinrich2015TheRO}, we propose a more data-driven approach: predicting an individual's intelligibility score for unseen audio based on their known intelligibility scores from other audio samples. Theoretically, if we had access to an individual's intelligibility score data across an infinite number of audio samples, predicting the intelligibility score would essentially become a lookup table problem. While obtaining infinite data is not feasible, we believe that with a sufficient amount of intelligibility score data from known audio, we can make reliable intelligibility score prediction.

\begin{figure}
    \centering
    \includegraphics[width=1\linewidth]{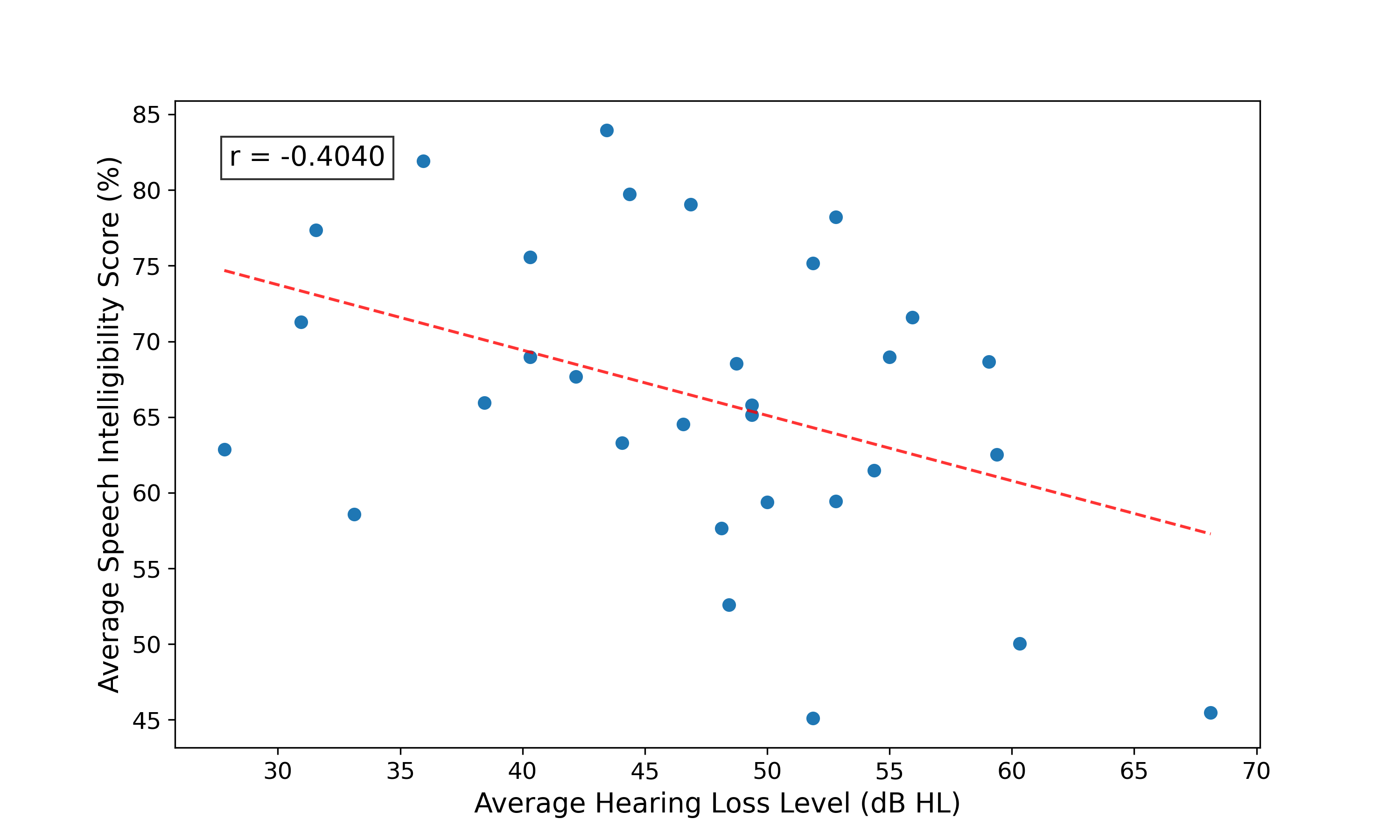}
    \caption{The listener's audiogram cannot fully manifest his/her speech intelligibility score. Although in general speech intelligibility score decreases with an increasing hearing loss level, large variances can occur between individuals with similar audiograms. Each point represents one listener. The dotted line is the linear regression curve and $r$ is the Pearson Correlation Coefficient.}
    \label{fig:enter-label}
\end{figure}

In this work, we introduce SSIP (\textbf{S}upport \textbf{S}ample-Based \textbf{I}ntelligibility \textbf{P}rediction), a novel framework that leverages an individual’s past intelligibility scores on known audio to predict the intelligibility of new audio clips. We draw inspiration from few-shot learning \cite{1597116, 10.5555/3157382.3157504, 10.5555/3294996.3295163}, where we refer to the (audio, score) pairs with known intelligibility scores as support samples, and the audio samples for which we need to predict intelligibility scores as query samples. To solve the problem under this framework, we proposed an innovative model architecture called SSIPNet. SSIPNet comprises two core components: (1) a speech foundation model-based intelligibility feature extractor that captures phonetic, semantic and perceptual cues; and (2) a predictor that aggregates an arbitrary number of (audio, score) pairs into a personalized listener embedding, from which the intelligibility score is predicted. Unlike standard audiogram-based methods, SSIPNet implicitly models higher-order perceptual and cognitive factors that pure tone audiometry fails to capture, offering enhanced personalization. Our experiments show that, even with a relatively small set of known intelligibility scores, our method outperforms state-of-the-art audiogram-based baselines on the CPC dataset.
 
In conclusion, our work presents \textbf{SSIP}, a novel approach to personalized intelligibility prediction, illustrating how a purely data-driven method, without relying on human-defined features, can surpass traditional audiogram-based methods. By demonstrating the feasibility and benefits of extracting highly granular, listener-specific embeddings, we aim to inspire future research focused on more refined data collection and model design that more effectively captures individual variability. Ultimately, this shift toward personalized, data-driven approaches has the potential to substantially improve the experience of hearing aid users in real-world listening environments, complementing and advancing both speech enhancement and other hearing aid algorithm developments.

The rest of the paper is organized as follows. Section 2 provides a detailed description of our proposed model. In Section 3, we present the experimental setup and results. Section 4 discusses the conclusion and outlines future directions.

\section{Method}
\subsection{Audio-audiogram decorrelation}
We observed a strong correlation between the listener's audiogram and the RMS level of the audios presented to that listener on the CPC dataset, as shown in Figure 2. This correlation is expected, as the audios are generated by different enhancement systems, each aimed at providing optimal compensation for individual listeners \cite{barker22_interspeech, 10446441}. However, since our goal is to assess whether using (audio, score) pairs alone can yield satisfactory prediction results, we aim to eliminate this correlation to exclude audiogram information.

\begin{figure}
    \centering
    \includegraphics[width=0.8\linewidth]{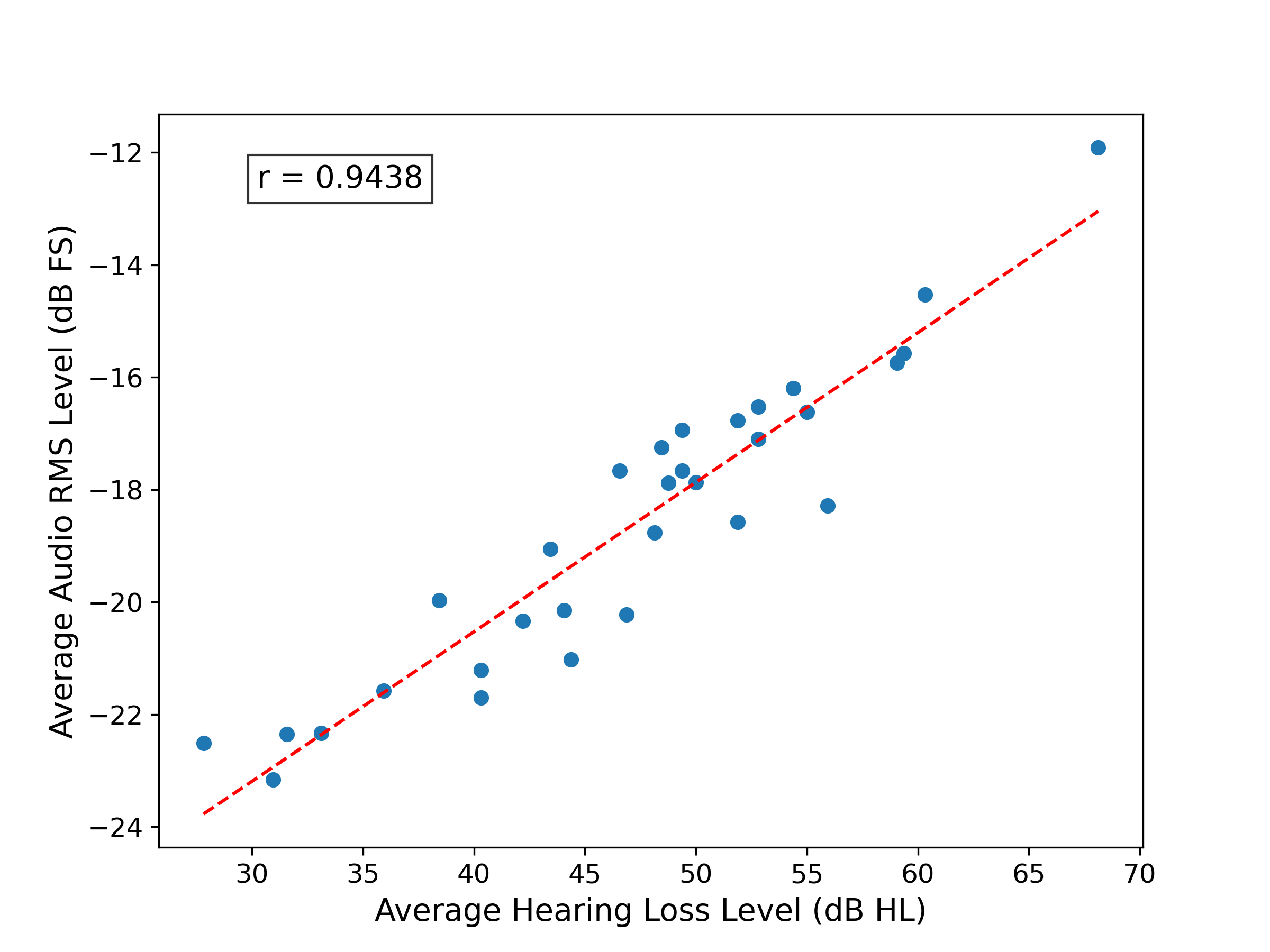}
    \caption{The levels of audios played to each listener have a strong correlation with his/her audiogram on the CPC dataset. Each point represents one listener. The dotted line is the linear regression curve and $r$ is the Pearson Correlation Coefficient.}
    \label{fig:enter-label}
\end{figure}

To address this, we draw on the audio level-intelligibility relationship curves from \cite{Hood01011971}, assuming that all listeners in the CPC dataset have conductive hearing loss. We normalize all audio to 65 dB SPL and use these curves to calibrate the corresponding ground truth intelligibility scores. The calibration curve we applied is shown in Figure 3. As a result of normalizing all audios to the same RMS level, the model is less likely to access any audiogram-related information that might have been encoded in the original audio levels. This removes the potential information leakage introduced by the correlation between the audiogram and the RMS level, forcing the model to predict intelligibility solely from existing scores. 

\begin{figure}
    \centering
    \includegraphics[width=1.0\linewidth]{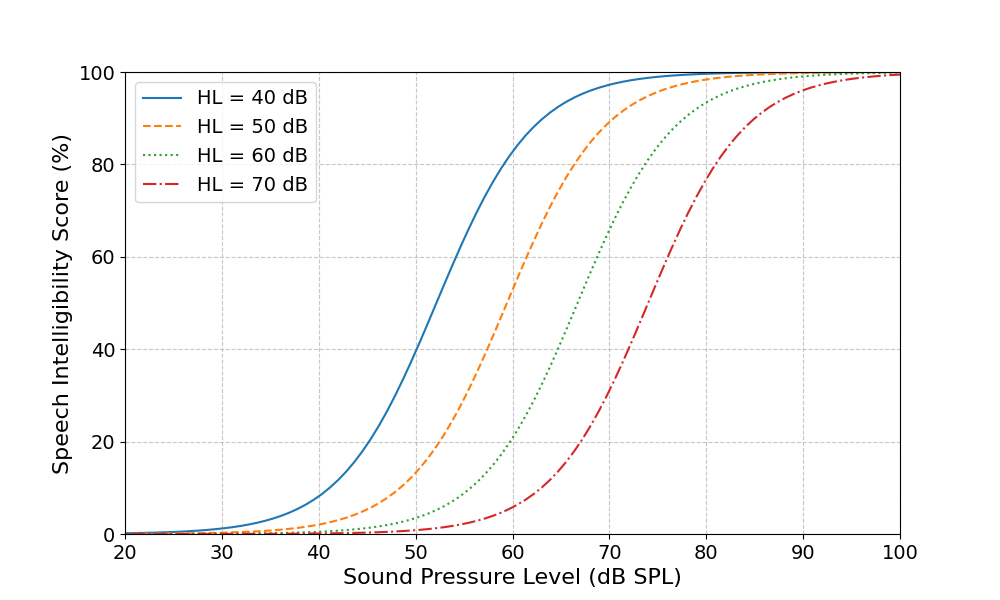}
    \caption{The audio level-intelligibility relationship curves are applied to obtain the calibrated intelligibility scores for audios at 65 dB SPL. Here we only show the curves at 4 different average hearing loss levels. Let the original audio level be $L_0$, with its corresponding intelligibility score $S_0$ from the CPC dataset. If the audio level is then adjusted to $L_1$, the calibrated intelligibility score becomes: $S_1 = S_0 + C_{HL}(L_1) - C_{HL}(L_0)$, where $HL$ represents one's average hearing loss level at 500, 1000, and 2000Hz and $C_{HL}$ is the corresponding audio level-intelligibility function.}
    \label{fig:enter-label}
\end{figure}

\subsection{Model}
Our model is composed of two key components: the Feature Extraction Module (FEM) and the Support-Based Prediction Module (SPM), as illustrated in Figure 4. For the FEM, we leverage the approach from the CPC2 non-intrusive track champion \cite{10447907}, where a fixed-length embedding is extracted from the encoder layers of a speech foundation model. In the SPM, we aggregate information from the embeddings of the support (audio, score) pairs, which is then combined with the query embedding to make the final speech intelligibility score prediction.

\subsubsection{Feature extraction module}
Previous studies have shown that features extracted from speech foundation models can effectively predict speech intelligibility \cite{10447907, zezario24_interspeech, best24_interspeech}. In this work, we directly adopt the same architecture as in \cite{10447907}. A pre-trained frozen backbone is used to extract meaningful representations from the input audio. The outputs from all encoder layers of the backbone network yield $L$ features of shape $(t \times d_{1})$, where $L$ is the number of encoder layers, $t$ and $d_{1}$ represent the time length and number of channels of the feature, respectively. Next, these features undergo temporal pooling and a temporal transformer layer, producing $L$ time-squeezed features of shape $(1 \times d_{2})$, where $d_{2}$ is the new number of channels. Unlike \cite{10447907}, which incorporates a linear-projected audiogram feature, we instead concatenate a linear-projected score feature along the layer dimension. For support audio samples, the input to this linear projection is their ground truth intelligibility score. For query audios samples, since their intelligibility scores are unknown, we set them to -1. Finally, these $L+1$ features of shape $(1 \times d_{2})$ are passed through a layer transformer followed by a layer pooling operation to generate a single $(1 \times d_{2})$ embedding.

We follow \cite{10447907} for hyperparameter settings and we use the large-size Whisper model v3 as the speech foundation model. In our setup, we set $L=32$, $d_{1}=1280$ and $d_{2}=384$.

\begin{figure*}
    \centering
    \includegraphics[width=1.0\linewidth]{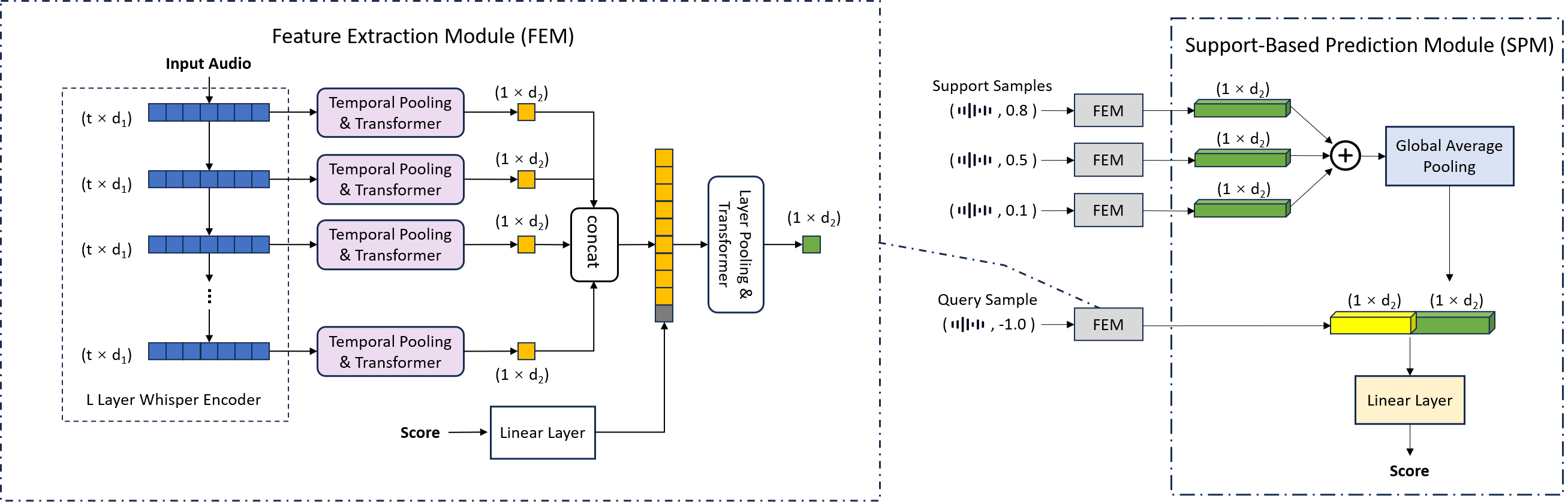}
    \caption{Our SSIPNet model structure. It follows a pipeline of sample-level embedding learning, listener-level feature aggregation, and final intelligibility score prediction. Module blocks with the same color indicate shared weights. The left side represents our speech foundation model-based feature extraction module, where the channel dimension has been omitted for simplicity. The right side illustrates our Support-Based Prediction Module, which uncovers relationships between existing scores to help make prediction.}
    \label{fig:enter-label}
\end{figure*}

\subsubsection{Support-based prediction module}
The FEM is designed to extract high-dimensional audio-score features for single audio prediction. To utilize information from multiple support (audio, score) pairs belonging to the same listener, we introduce the Support-Based Prediction Module (SPM). The SPM aggregates embeddings extracted from the support (audio, score) pairs and integrates them with the query embedding to assist in predicting the intelligibility score of the query audio. Theoretically, this aggregation operation needs to handle set-type inputs, where the permutation-invariant property corresponds to the fact that our support samples have no inherent order, and the collection-of-elements property captures the fact that multiple support samples belong to the same listener. Here, element-wise global average pooling (GAP) \cite{lin2013network}, which is simple yet satisfies the requirement, is applied to the support embeddings, which are then concatenated channel-wise with the query embedding. Finally, a linear layer projects the concatenated embeddings into the final prediction score for the query audio.

\subsection{Training and inference mechanism}
Unlike standard speech intelligibility score prediction, where each batch consists of multiple independent (audio, audiogram) pairs, in our setup, each batch is structured by listener, with each listener having multiple (audio, score) pairs as support samples and several query audio samples for prediction.

\section{Experiments and results}
\subsection{Experimental setup}

\subsubsection{Data}
For our experiments, we used the Clarity Prediction Challenge (CPC) dataset \cite{barker22_interspeech, 10446441}, which consists of 13126 distinct audio signals with experimentally-acquired intelligibility scores. These audio samples are generated by 18 enhancement systems and subsequently presented to 27 hearing-impaired listeners. For training, validation and testing, we followed the three-fold  strategy as in \cite{10446441}. However, we did not directly use the existing test set, as it contained too few audio samples per listener, making it unsuitable for evaluating our method with a relatively large number of support audio samples. Instead, we restructured the training, validation, and test sets and we ensured that listeners in the training, validation, and test sets did not overlap. Specifically, the numbers of samples and listeners in the training, validation, and test set are shown in Table 1.

\begin{table}[ht]
  \caption{Number of samples and listeners in the training, validation, and test sets. The sample numbers correspond to the three folds, while the number of listeners remains consistent across all folds.}
  \label{tab:example}
  \centering
  \begin{tabular}{c c  c}
    \toprule
    \textbf{Type} & \textbf{Sample Num} & \textbf{Listener Num} \\
    \midrule
    Training           & $7925/7461/7220$ & 23 \\
    Validation            & $674/674/676$ & 3\\
    Test            & $2455/2903/3166$ & 5\\
    \bottomrule
  \end{tabular}
\end{table}

\subsubsection{Baseline}
We compare our method against the standard approach based on (audio, audiogram) pairs. As our baseline, we use the E011 method from the CPC2 \cite{10447907}, which is the champion of the non-intrusive track. We have re-implemented this method and evaluated it on our test set. 

\subsubsection{Evaluation metric}
To evaluate the performance of predictors in our experiments, we use two metrics.
The first is the root mean square error (RMSE), defined as:
\begin{align}
  E(\hat{y}, y) &= \sqrt{\frac{1}{N}\sum_{i=1}^N{(\hat{y}_i - {y}_i)^2}},
  \label{equation:eq2}
\end{align}
where $\hat{y}$ represents the predicted intelligibility score, $y$
is the ground truth intelligibility score, and $N$ is the total number of query audio samples.

The second metric is the normalized Pearson correlation coefficient (NCC), given by:
\begin{align}
  R(\hat{y}, y) &= \frac{\sum_{i=1}^N{(\hat{y}_i-\overline{\hat{y}})(y_i-\overline{y})}}{\sqrt{\sum_{i=1}^N{(\hat{y}_i-\overline{\hat{y}})^2}}\sqrt{\sum_{i=1}^N{(y_i-\overline{y})^2}}},
  \label{equation:eq3}
\end{align}
where $\hat{y}$ is the predicted intelligibility score, $y$
is the ground truth intelligibility score, $\overline{\hat{y}}$ is the mean predicted intelligibility score, and $\overline{y}$ is the mean ground truth intelligibility score. N is the total number of query audio samples.

\subsubsection{Training and testing details}
We trained all the models for 200 epochs using Huber loss \cite{10.1214/aoms/1177703732} with the Adam optimizer \cite{DBLP:journals/corr/KingmaB14}. The learning rate, $\beta_{1}$, and $\beta_{2}$ were set to $3\mathrm{e}{-5}$, $0.9$ and $0.98$, respectively. We also employed a cosine annealing scheduler \cite{DBLP:conf/iclr/LoshchilovH17} with a linear warmup \cite{goyal2017accurate} for the first 10 epochs. The start factor was set to $0.1$. The batch size was set to 128, with a fixed number of support samples in each batch. The remaining samples in each batch were query samples. Both support and query samples were randomly selected from the same listener during each training iteration. During testing, the support and query samples were fixed. The model weights that achieved the best performance on the validation set were used for the final evaluation on the test set, and we recorded the output scores.

\subsection{Effectiveness of support (audio, score) pairs}
Table 2 and Table 3 present the results of our method alongside the baseline methods on the test set. We used $64$ support (audio, score) pairs and averaged the results over three folds. The results demonstrate that, by incorporating additional (audio, score) pairs and leveraging a well-structured model, our method outperforms the baseline, which is already a strong one. We attribute this improvement to the inclusion of extra (audio, score) pairs, which enable the model to capture more nuanced hearing information than is available from audiograms alone. This, in turn, leads to more personalized and accurate intelligibility score predictions.

\begin{table}[ht]
  \caption{RMSE values for our method and the baseline, showing results across three test sets and the average value. Lower is better.}
  \label{tab:example}
  \centering
  \begin{tabular}{ c  c}
    \toprule
    \textbf{Method} & \textbf{RMSE} \\
    \midrule
    SSIPNet (Ours)           & $\textbf{22.693}/\textbf{25.176}/\textbf{23.809}/\textbf{23.430}$  \\
    Baseline            & $23.450$/$26.283/28.746/26.160$ \\
    \bottomrule
  \end{tabular}
\end{table}

\begin{table}[ht]
  \caption{NCC values for our method and the baseline, showing results across three test sets and the average value. Higher is better.}
  \label{tab:example}
  \centering
  \begin{tabular}{ c  c}
    \toprule
    \textbf{Method} & \textbf{NCC} \\
    \midrule
    SSIPNet (Ours)           & $\textbf{0.816}/0.787/\textbf{0.809}/\textbf{0.811}$  \\
    Baseline            &  $0.803/\textbf{0.809}/0.770/0.794$ \\
    \bottomrule
  \end{tabular}
\end{table}

\subsection{Influence of number of support (audio, score) pairs}
In addition to the previous experiment, an important question is how the number of additional support (audio, score) pairs affects prediction performance. To explore this, we examine the impact of varying the number of support (audio, score) pairs, ranging from 1 to 64, as shown in Figure 5. From the figure, we observe the following: (1) A noticeable improvement in performance occurs when the number of support (audio, score) pairs is at least 4. Beyond this point, further increasing the number of pairs does not lead to a significant boost in performance. (2) The best performance is achieved when the number of support (audio, score) pairs is 16. We hypothesize that, for a fixed model size, introducing too many support (audio, score) pairs may exceed the model’s capacity, leading to diminishing returns and potential confusion. (3) Even with as few as one support (audio, score) pair, the model performs comparably to the standard (audio, audiogram) setup. This can be attributed to the fact that, despite the (audio, audiogram) decorrelation process we applied, some correlation remains. Furthermore, the audiogram distribution in the CPC dataset is relatively concentrated, so the audiogram alone is not very informative. In contrast, even a single (audio, score) pair can reveal valuable differences in speech recognition ability beyond what the audiogram provides.

\begin{figure}
    \centering
    \includegraphics[width=1.0\linewidth]{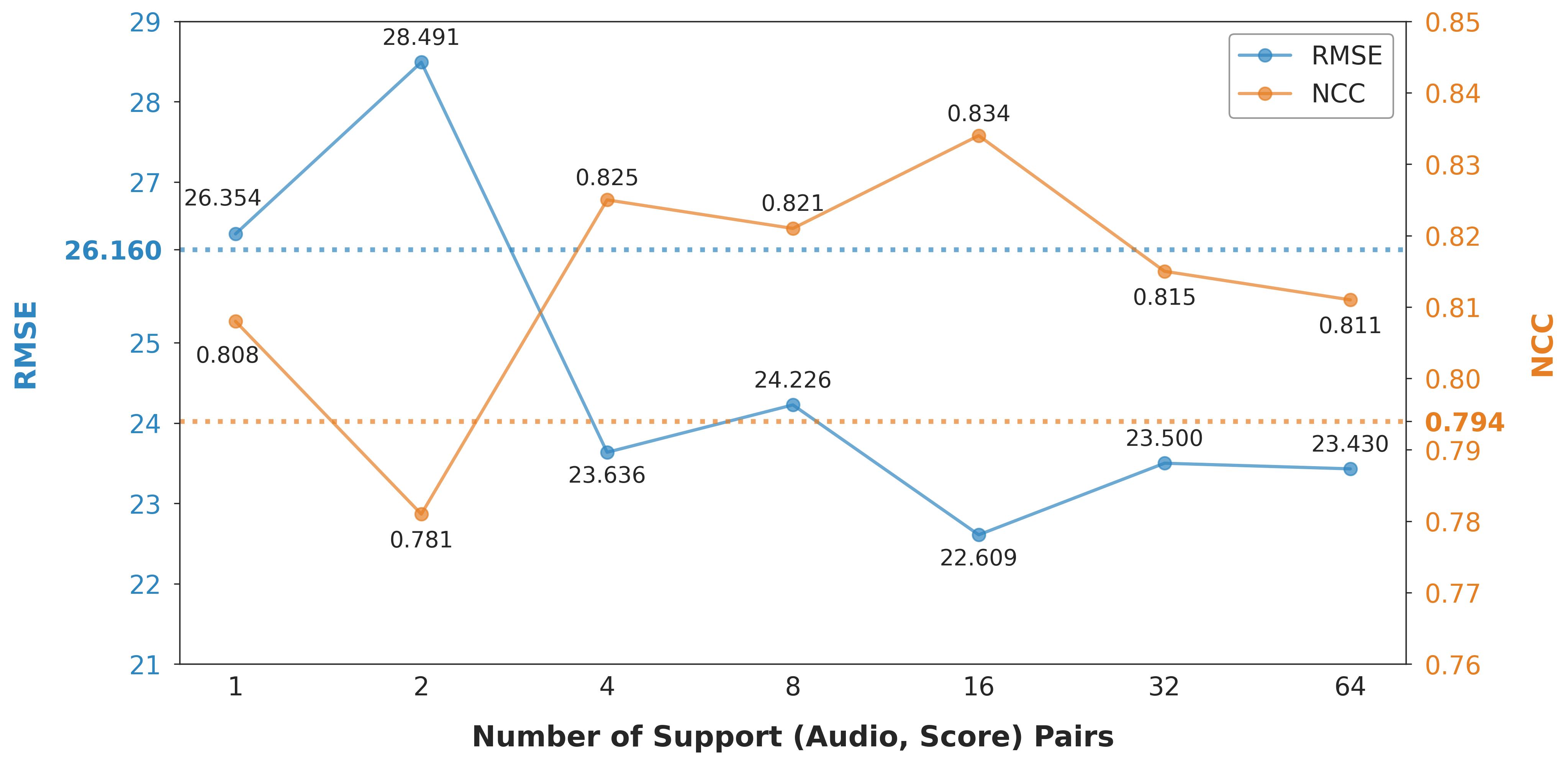}
    \caption{Relationship between prediction performance and the number of support (audio, score) pairs. The dotted horizontal lines indicate the audiogram-based baseline. The numbers represent the average results across the three folds.}
    \label{fig:enter-label}
\end{figure}

\section{Conclusion}
This paper introduces SSIP, a method for personalized speech intelligibility prediction that generalizes intelligibility scores from existing audio samples to unseen audio. Our experiments demonstrate that, with a carefully designed model architecture, methods leveraging a relatively small number of support (audio, score) pairs can outperform standard audiogram-based approaches on the CPC dataset. This highlights the potential for more granular modeling of speech recognition ability, without the need for complex additional features. From a broader perspective, we consider this problem as part of the few-shot regression domain, which currently lacks as mature solutions as those for few-shot classification. Consequently, we believe that focusing on model architectures and loss functions specifically designed for few-shot regression tasks could be a promising avenue for future research. Further investigations could explore datasets with greater variability in hearing abilities to evaluate the robustness of the method. Additionally, future work could examine which types of intelligibility-labeled audio are most effective in building a more accurate hearing profile for listeners, thereby enhancing prediction performance for unseen audio with minimal support samples.

\bibliographystyle{IEEEtran}
\bibliography{export}

\end{document}